\documentclass[twocolumn]{aastex63}
\usepackage{graphicx}
\usepackage{subfigure}
\usepackage{color, hyperref, epsfig}
\usepackage{apjfonts, natbib}
\usepackage{appendix}
\usepackage{float}
\usepackage{bm}

\newcommand{\ha}{H\ensuremath{\alpha}}
\newcommand{\hb}{H\ensuremath{\beta}}

\newcommand{\sdss}{\emph{SDSS}}

\newcommand{\hab}{H\ensuremath{\alpha,_B}}

\newcommand{\nii}{[N\,{\footnotesize II}]}
\newcommand{\sii}{[S\,{\footnotesize II}]}

\newcommand{\heii}{He\,{\footnotesize II} }
\newcommand{\oiii}{[O\,{\footnotesize III}]}

\newcommand{\kms}{\ensuremath{\mathrm{km~s^{-1}}}}
\newcommand{\mbh}{\ensuremath{M_\mathrm{BH}}}
\newcommand{\wise}{WISE}

\shorttitle{MIRONG Sample II}
\shortauthors{Wang et al.}

\begin{document}

\title{Mid-InfraRed Outbursts in Nearby Galaxies (MIRONG). II. Optical Spectroscopic Follow-up}

\author[0000-0003-4225-5442]{Yibo~Wang}
\affiliation{CAS Key Laboratory for Research in Galaxies and Cosmology,
Department of Astronomy, University of Science and Technology of China, 
Hefei, 230026, China; wybustc@mail.ustc.edu.cn,jnac@ustc.edu.cn,shengzf@ustc.edu.cn,twang@ustc.edu.cn}
\affiliation{School of Astronomy and Space Sciences,
University of Science and Technology of China, Hefei, 230026, China}

\author[0000-0002-7152-3621]{Ning Jiang}
\affiliation{CAS Key Laboratory for Research in Galaxies and Cosmology,
Department of Astronomy, University of Science and Technology of China, 
Hefei, 230026, China; wybustc@mail.ustc.edu.cn,jnac@ustc.edu.cn,shengzf@ustc.edu.cn,twang@ustc.edu.cn}
\affiliation{School of Astronomy and Space Sciences,
University of Science and Technology of China, Hefei, 230026, China}

\author[0000-0002-1517-6792]{Tinggui Wang}
\affiliation{CAS Key Laboratory for Research in Galaxies and Cosmology,
Department of Astronomy, University of Science and Technology of China, 
Hefei, 230026, China; wybustc@mail.ustc.edu.cn,jnac@ustc.edu.cn,shengzf@ustc.edu.cn,twang@ustc.edu.cn}
\affiliation{School of Astronomy and Space Sciences,
University of Science and Technology of China, Hefei, 230026, China}

\author[0000-0003-1710-9339]{Lin~Yan}
\affiliation{Caltech Optical Observatories, California Institute of Technology, Pasadena, CA 91125, USA; lyan@caltech.edu}

\author[0000-0001-6938-8670]{Zhenfeng~Sheng}
\affiliation{CAS Key Laboratory for Research in Galaxies and Cosmology,
Department of Astronomy, University of Science and Technology of China, 
Hefei, 230026, China; wybustc@mail.ustc.edu.cn,jnac@ustc.edu.cn,shengzf@ustc.edu.cn,twang@ustc.edu.cn}
\affiliation{School of Astronomy and Space Sciences,
University of Science and Technology of China, Hefei, 230026, China}\

\author[0000-0002-4757-8622]{Liming Dou}
\affiliation{Department of Astronomy, Guangzhou University, Guangzhou 510006, China; doulm@gzhu.edu.cn}

\author[0000-0003-4651-8510]{Jiani Ding}
\affiliation{Department of Astronomy and Astrophysics, UCO/Lick Observatory, University of California, 1156 High Street, Santa Cruz, CA 95064, USA}

\author[0000-0001-8467-6478]{Zheng Cai}
\affiliation{Department of Astronomy, Tsinghua University, Beijing 100084, People's Republic of China}

\author[0000-0002-7223-5840]{Luming~Sun}
\affiliation{Department of Physics, Anhui Normal University, Wuhu, Anhui, 241000, People's Republic of China}

\author[0000-0003-4975-2333]{Chenwei~Yang}
\affiliation{Polar Research Institute of China, 451 Jinqiao Road, Shanghai, 200136, People's Republic of China}

\author[0000-0002-7020-4290]{Xinwen~Shu}
\affiliation{Department of Physics, Anhui Normal University, Wuhu, Anhui, 241000, People's Republic of China}


\submitjournal{\apjs}
\received{ 2021 August 31}
\revised{2021 October 13}
\accepted{ 2021 October 25}

\begin{abstract}
Infrared echo has proven to be an effective means to discover transient accretion events of supermassive black holes (SMBHs), such as 
tidal disruption events (TDEs) and changing-look active galactic nuclei (AGNs), in dusty circumnuclear environments. To explore the 
dusty populations of SMBH transient events, we have constructed a large sample of Mid-infrared Outbursts in Nearby Galaxies (MIRONG) 
and performed multiwavelength observations. Here we present the results of multiepoch spectroscopic follow-up observations of a subsample 
of 54 objects spanning a time scale of 4 yr. Emission-line variability was detected in 22 of them with either emergence or enhancement 
of broad Balmer emission lines in comparison with pre-outburst spectra. Coronal lines, $\rm He\,\textsc{ii}\lambda4686$ and Bowen line 
$\rm N\,\textsc{iii}\lambda4640$ appeared in the spectra of nine,seven and two sources, respectively.  These results suggest that MIRONG 
is a mixed bag of different transient sources. We have tentatively classified them into different subclass according to their spectral 
evolution and light curves. Two sources have been in a steady high broad \ha\ flux up to the latest observation and might be turn-on AGNs. 
Broad lines faded out in the remaining sources, indicating a transient ionizing source ignited by TDE or sporadic gas accretion. 
Thirty-one sources do not show noticeable spectral change with respect to their pre-outburst spectra. They have a statistically redder MIR 
color and lower MIR luminosity of the outbursts,which are consistent with heavily obscured events.
\end{abstract}


\section{Introduction}
\label{intro}

Time-domain astronomy, which is promoted by wide-field, deep and fast 
surveys, has experienced great progress recently and resulted in the 
explosive growth of the discovery speed of various transients. Among them, 
the nuclear transients associated with supermassive black holes (SMBHs) 
have gradually become a regular class of targets in the past decade though their nature remains to be explored in many cases 
(e.g., \citealt{Kankare2017,Trakhtenbrot2019,Hinkle2021,Malyali2021}). 

SMBHs are ubiquitous in the centers of massive galaxies. Most of them are inactive in the local universe (\citealt{KH2013}). 
Occasionally, a dormant SMBH undergoes a temporary accretion phase when a star in the nucleus wanders within the tidal radius and 
gets torn apart. Part of the stellar debris will be accreted by the SMBH, producing a luminous flash of electromagnetic radiation 
peaked at soft X-ray to ultraviolet (UV) bands. The process is called a tidal disruption event (TDE; \citealt{Rees1988,Evans1989,Phinney1989}). 

Though predicted in the 1970s, the discovery of TDEs was not realized until the late 1990s  (\citealt{Bade1996}) because their rarity 
makes the discovery rather challenging. First, the duration of a TDE is short, which normally rises to a peak in $\sim$one month and 
then decays in months to years. Second, the event rate is as low as $10^{-4}-10^{-5}$/galaxy/year (\citealt{Wang2004,Stone2016}). 
The accumulated number of TDEs to date remains in two digits. Nevertheless large optical surveys, such as the Zwicky Transient Facility (ZTF), 
have greatly advanced the searching efficiency in the past few years (see review by \citealt{Gezari2021}), and will eventually bring the 
field into the era of population study (\citealt{vV2021}). TDEs offer us a unique opportunity to probe SMBHs in normal galaxies 
(\citealt{Lu2017,Mockler2019,Pasham2019}) and the life cycle of accretion disks (e.g., \citealt{Wevers2019}) as well as jets 
(e.g., \citealt{Bloom2011,Burrows2011,Mattila2018,TDEradio}).

In addition to TDEs, SMBHs may also experience sporadic gas accretion due to instability of the accretion flow or stochastic infall 
of cold gas clouds, such as G1/2 in the Galactic center. In some active galactic nuclei (AGNs), broad emission lines appear/disappear 
in response to the dramatic change of the central ionizing continuum on time scales from a few months to a few years. They are dubbed 
as "changing-look" AGNs (e.g., \citealt{Shappee2014,LaMassa2015,MacLeod2016,Yang2018,Guo2019,Sheng2020}). In the most extreme cases, 
a quiescent galaxy is transformed to an AGN or vice versa, called "turn-on" (\citealt{Gezari2017,Frederick2019,Yan2019}) or "turn-off" 
AGNs (e.g., \citealt{Runnoe2016}). The dramatic change of their accretion rate has already posed a serious challenge to canonical 
accretion disk theories. Moreover, the event rate statistics of changing-look AGNs may advance our knowledge of the AGN duty cycle, 
which might impact the evolution of their host galaxies via feedback. 

Given the importance of TDEs and changing-look AGNs, a systematic search for them is a pressing need. The search hitherto has mainly 
focused on the optical band and yielded a mounting number of TDEs (e.g., \citealt{vV2020,vV2021}) and turn-on AGNs (e.g., \citealt{Frederick2019}) 
discovered in real time, which allows prompt follow-up observations. However, the optical search is unavoidably affected by dust 
extinction and thus may overlook events in dusty environments. For instance, the low dust-covering factor of optical TDEs suggests that 
the optical band might be merely effective in unveiling TDEs in very dust-poor environments (\citealt{Jiang2021b}). 
Fortunately, obscured events may expose themselves as reprocessed infrared (IR) emission by dust in the vicinity of SMBHs (\citealt{Lu2016}), 
namely IR echoes of TDEs (\citealt{Dou2016,Jiang2016,vV2016,vV2021b}) and changing-look AGNs (\citealt{Sheng2017}), which are relatively 
transparent in the mid-infrared (MIR) to a moderately thick obscurer. Systematic searches in the IR have already led to successful 
discoveries of completely obscured TDEs (e.g., \citealt{Mattila2018,Kool2020}). The other potential use of an IR search is its ability 
to reveal transients that are optically weak but luminous in higher-energy bands (e.g., UV and X-rays). Actually, classic TDE models 
predict emissions from accretion exactly in such wavelength regimes. 

Encouraged by the powerful IR echoes (e.g., \citealt{Wang2018,Sheng2020}), 
we have recently carried out a systematic search of ongoing outbursts 
in the MIR band using archival Wide-field Infrared Survey Explorer (\wise) light curves. 
We obtained a sample of 137 MIR outbursts in nearby galaxies (MIRONG) 
at $z<0.35$, which showed a sudden brightening of $>0.5$ mag 
in at least one \wise\ band (\citealt{Jiang2021a}, hereafter Paper I). 
The Majority of these MIRONG do not have any known corresponding optical flares, 
indicating dust extinction or intrinsic weakness in the optical band. 
Their MIR luminosities are markedly brighter than those of known supernovae (SNe) and their 
physical locations are very close to the galactic center (median $<0\arcsec.1$), 
suggesting strongly that they are dust echoes of transient accretion events 
onto SMBHs. Multiwavelength follow-up observations have been undertaken 
since 2017 to investigate the nature of the newly unveiled MIRONG. 
In particular, their optical spectroscopic observations are a key part of 
the monitoring and can provide the most direct imprint of the outburst 
by probing the variability of characteristic emission lines.  

This paper is the second one in the series of MIRONG and 
is dedicated specifically to the spectral evolution of MIRONG. 
We assume a cosmology with $H_{0} =70$ km~s$^{-1}$~Mpc$^{-1}$, 
$\Omega_{m} = 0.3$, and $\Omega_{\Lambda} = 0.7$.

\section{Sample Description and Observation Strategy}

The spectroscopic sample was drawn from the parent MIRONG sample as described in 
Paper~I.  In order to capture potential spectral signals in a timely manner for 
rapidly evolving transients, such as TDEs, we select targets that are 
still in the early stage of the outburst for spectroscopic observations.
. 
Our observing campaign started in 2017 April, and a total of 54 
objects have been observed up to 2021 March. The median redshift 
is 0.09, which is similar to the parent sample.

The observations were carried out with either the Double spectrograph (DBSP) mounted on the Hale 200 inch telescope 
at Palomar observatory (\citealt{Oke1982}) or the Kast DBSP on the Shane telescope at Lick observatory (\citealt{Miller1993}). 
We adopt the D55 dichroic for Hale 200 inch telescope, which splits the incoming light  into separate red and blue channels 
at $5500$~\AA\ for two different gratings, with a blue grism  of 600 lines per mm blazed at 3780~\AA\ and a red grism of 316 lines 
per mm  blazed at 7150~\AA\ . The slit width of 1\arcsec, 1.\arcsec5 or 2\arcsec\ was used depending on the weather. Such setup 
achieves a resolution of 5.23-10.46~\AA\ in the red channel and 2.76-5.32~\AA\ in the blue channel with a full spectral wavelength 
coverage of about 3100-10500~\AA . With a proper exposure time chosen for each source, the median signal-to-noise ratios (S/N) are 
14.2 $\rm pixel^{-1}$ and 25.6 $\rm pixel^{-1}$, respectively, for the blue- and red-end DBSP spectra.  The Shane telescope was configured with 
a D57 dichroic, blue grism of 600 lines per mm blazed at 4130~\AA, red grating of 600 lines per mm blazed at 7500 \AA\ and slit widths 
of 1\arcsec\ or 1.\arcsec5 , which result in a resolution of 2.4-3.6~\AA\ on the blue end and 3.0-4.5~\AA\ on the red end and a full 
spectral wavelength coverage of about 3000-10000~\AA\ . Slightly lower median S/N of 6.1 $\rm pixel^{-1}$ (blue) and 11.8 $\rm pixel^{-1}$ (red) 
were approached. Besides, there are also two spectra acquired for J1115+0544 from Keck and LAMOST (\citealt{Yan2019}).
Details on the exposure time and slit selection for every single observation could be found in the Appendix \ref{appendixA}.

All spectra were reduced according to the standard reduction procedures for a long-slit spectrum. They were flux calibrated with 
the standard stars observed in the same night. 

\section{Data Analysis and Results} 

The main goal of this work is to identify the spectral features accompanying the MIR outburst by comparing the spectra taken before 
and after the MIR outburst, and their subsequent evolution. Although variations of prominent emission lines are obvious in many cases, 
a thorough analysis is essential to acquire robust detection and measurements of weak emission lines and for the entire sample. 

\subsection{Subtraction of the Continuum}

Aiming at detecting potential characteristic emission-line variation, we began our analysis with subtraction of continuum. 
These spectra are first corrected for Galactic extinction using the extinction map of \citet{Schdustmap} and the extinction 
curve of \citet{Fitz1999}.

\begin{figure*}
\figurenum{1}
\centering
\begin{minipage}{0.9\textwidth}
\centering{\includegraphics[angle=0,width=1.0\textwidth]{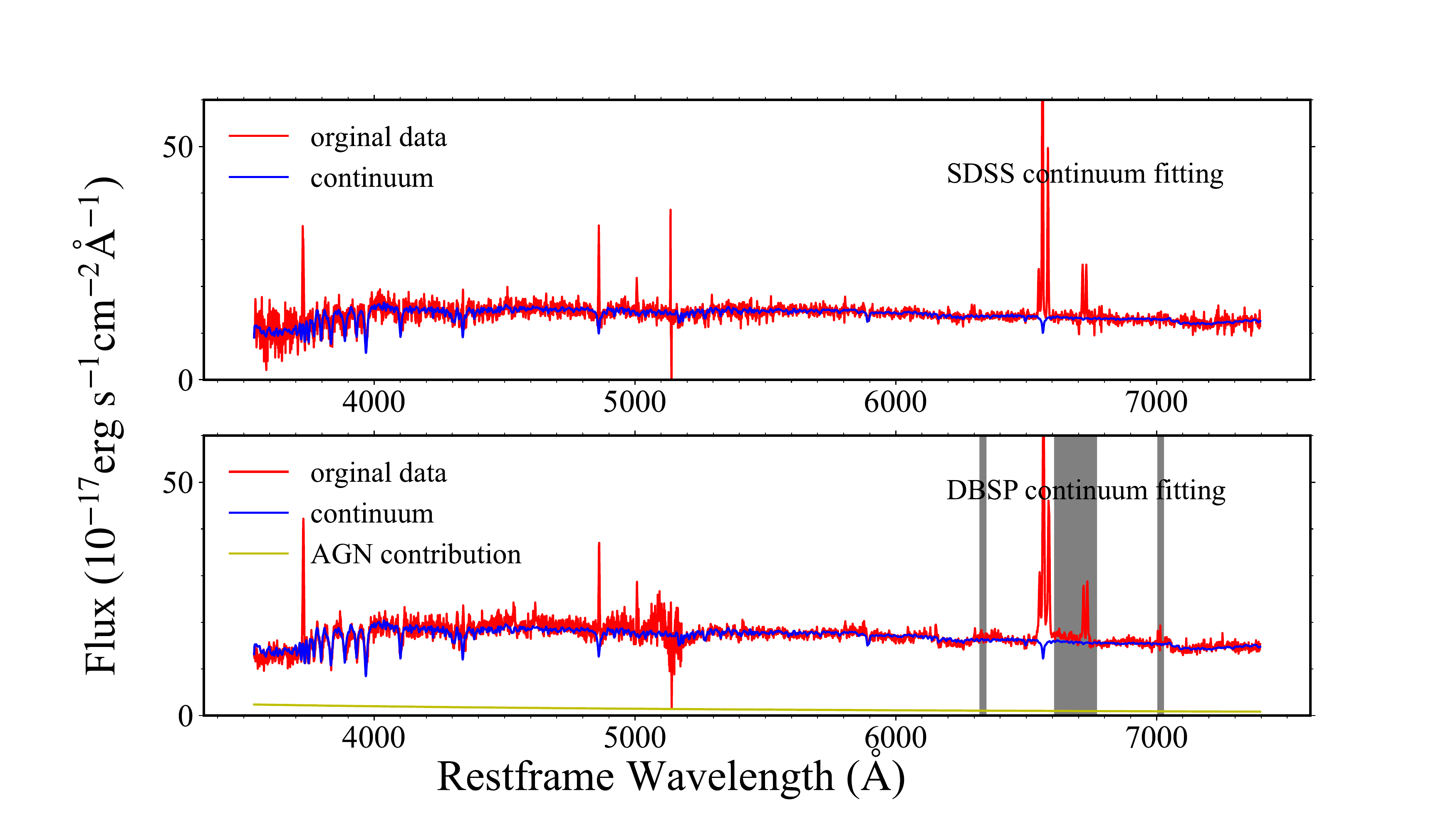}}
\end{minipage}
\caption{We show the continuum fitting of the spectrum of J1647+3843
before (SDSS, top panel) and after (DBSP, bottom panel) an MIR outburst as an example.
The original SDSS spectrum is plotted in red and the fitted stellar continuum 
is shown in blue. For the DBSP spectrum taken after the outburst, an additional 
reddened power-law component (yellow) has been also included in the fitting. The shadow 
regions show the prominent telluric absorption windows, which were masked in the fitting procedure. 
\label{fig:continuumfit}}
\end{figure*}

Because the SDSS spectra are dominated by starlight, we fit them using the procedure {\tt PPXF} (\citealt{CE2004,Cappellari2017}) 
and MILES spectral library (\citealt{Vazdekis2010}). Nevertheless, the AGN continuum is likely not negligible for eight broad-line 
objects (\S~\ref{sec3.2}), while the {\tt PPXF} is a pure starlight component fitting code. So for these eight sources, we used 
the spectral decomposition results of Paper I instead. The fitting of J1647+3843 is shown as an example in Figure~\ref{fig:continuumfit}.

For the DBSP spectra observed after the MIR flare, an additional continuum component associated with the MIR outburst is possible. 
We thus also placed a reddened power-law continuum assuming an extinction curve of the Galactic dust (\citealt{Fitz1999}), as normally 
appears in broad-line AGNs, in the fitting. On the other hand, the shape of the starlight component remains exactly the same as that 
of the SDSS spectrum. Accordingly, we fit the continuum of DBSP spectra using the following formula: 
$$A*S(\lambda)+B*\lambda^\alpha*10^{-0.4*E(B-V)*k(\lambda)} \eqno (1)$$
$S(\lambda)$ is the starlight component of the same source obtained by applying the {\tt PPXF} procedure to the $\rm SDSS$ spectrum, and 
the power-law index $\alpha$ was set to be in the range between -2 and -1. The $k(\lambda)$ is the extinction curve and $E(B-V)$ 
varies from 0 to 2.5.

The fitting was done by Python code {\tt MPFIT}~\footnote{http://code.google.com/p/astrolibpy/source/browse/trunk/}
and the errors of the parameters were estimated from Monte Carlo simulations.
We show an example in the bottom panel of Figure~\ref{fig:continuumfit}.

\subsection{Emission-line Fitting and Detection Criteria}
\label{sec3.2}

\begin{figure*}
\figurenum{2}
\centering
\begin{minipage}{0.9\textwidth}
\centering{\includegraphics[angle=0,width=1.0\textwidth]{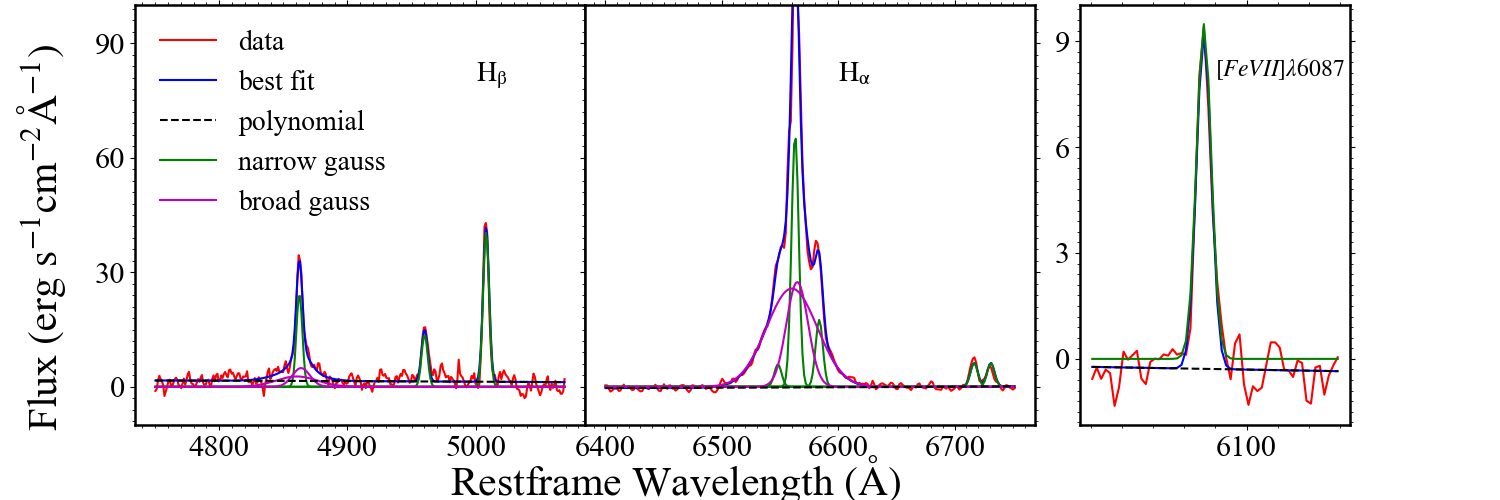}}
\end{minipage}
\caption{The left,middle,and right panel show the \hb  ,\ha  ,and $\rm [Fe\,\textsc{vii}]\lambda6087$ fitting 
results respectively. The red curves in the three panel represent the original emission lines data while 
the blue curves represent the best-fitting model. Narrow and broad components are displayed with green and purple curves. 
The black dashed line models the residual local continuum, and only a few spectra's \ha\ and \hb\ fitting need this 
component as shown in the figure to improve the fitting. 
\label{fig:linefit}}
\end{figure*}

After subtracting the continuum, the emission-line spectrum is modeled with a combination of Gaussian functions in several segments 
to measure the line fluxes of various broad and narrow lines. The broad lines include $\rm H\alpha$, $\rm H\beta$, \heii and Bowen 
line N\,\textsc{iii}. The narrow lines contain \oiii\ ,\nii\ ,\sii\ , [Ne\,\textsc{iii}]. Each narrow line is modeled with one or two 
Gaussians, while broad $\rm H\alpha$ and $\rm H\beta$ are fitted simultaneously with  Gaussians of the same width and centroid in 
velocity. The number of Gaussians is determined according to the 95\% threshold of improvement with the F-test by adding one more 
Gaussian to the fit. We show an example of $\rm H\alpha$ and $\rm H\beta$ fitting results in the Figure \ref{fig:linefit}.

However, a reliable detection of broad \ha\ requires not only at least one broad component as suggested by the F-test 
but a $5\sigma$ criterion was also demanded. Accordingly, eight sources in our sample have reliable broad \ha\ detection in the 
SDSS spectra, which are taken before the MIR flare (see appendix \ref{appendixA}). For DBSP spectra, usually with multiple epochs, 
we have fitted and tried to detect broad \ha\ in each of them. If there is no broad component detected in $\rm{H\alpha}$, we take 
the flux at the 90\% confidence level as the upper limit which is derived from $\Delta\chi^2=2.7$ (\Citealt{Avni1976}). We will adopt 
broad \ha, which is usually the strongest, if exists, to represent the evolution of broad components. 

We consider seven common coronal lines, namely $\rm [Fe\,\textsc{vii}]\lambda3759$, $\rm [Fe\,\textsc{v}]\lambda4071$,
$\rm{[Fe\,\textsc{xiv}]\lambda5304}$, $\rm{[Fe\,\textsc{vii}]\lambda5722}$, $\rm{[Fe\,\textsc{vii}]\lambda6087}$, 
$\rm{[Fe\,\textsc{x}]\lambda6376}$ and $\rm{[Fe\,\textsc{xi}]\lambda7894}$. Each is fitted with one 
Gaussian of FWHM$<$1500\kms\ while a linear function has also been added to represent the local residual continuum (see Figure \ref{fig:linefit} for an example). An F-test is 
employed to assess whether the Gaussian component is necessary or not. A line is detected if its flux is above the 3$\sigma$ level. 
The presence of coronal lines is defined as the detection of three coronal lines or two lines with one above the $5\sigma$ level 
in the same epoch, or a coronal line in multiple epochs. According to the above rule, no coronal lines were present in all our SDSS 
spectra but they show up in the post-outburst spectra of nine objects.  

\subsection{Emission-Line Variations}
\label{sec3.3}

Next we begin to check the emission-line variability between DBSP and SDSS 
spectra. Obviously, if broad \ha\ is reliably detected in the SDSS spectrum, one has 
to measure the line flux to assess whether it is variable or not. This requires a 
reliable flux calibration. Although our spectra were already flux calibrated 
using a standard star, due to changes in weather conditions over the night, significant 
uncertainty could be introduced. With a reliable broad \ha\ indicating significant nuclei
activity, recalibration was done based on the normalization of the $\rm [O\,\textsc{iii}]\lambda5007$ flux,
which is assumed to be constant for the Seyfert galaxy over the monitoring timescale (\Citealt{Peterson2013}). 
The variability of broad \ha\ then was acquired by checking whether its flux at any epoch during the 
follow-up observations has changed by $3\sigma$ relative to the SDSS one or not. However, 
recalibration could not be implemented for J2215-0107, one of the eight sources with reliable broad \ha\ detection in SDSS spectra , 
due to the too-low spectra quality, and we just removed it without further discussion (hence, 53 left).
Significant variation of broad \ha\ was found in the left seven. Counting those with newly 
emerged broad \ha\ , 22 in our samples display variation of broad \ha\ and the appearance of coronal lines 
was in 9 of them.
According to whether or not the variability of \hab\ or coronal lines is detected, our sample can be 
classified into three subclass (see Table~\ref{variation-t}). It is interesting to note that there is 
no object in our sample that shows only variation in coronal lines.

Additionally, as the $\rm [O\,\textsc{iii}]\lambda5007$ flux of the seven with reliable 
broad \ha\ in SDSS spectra and other Seyfert 2 galaxies was expected to be constant in our multiple DBSP follow-up spectra 
and the same as that of SDSS, the scatter of the $\rm [O\,\textsc{iii}]\lambda5007$ flux 
normalized by the SDSS ones could thus be used to estimate the accuracy of the original flux calibration. 
The median value of the scatter is about 0.17 dex. When we focused on the long-term evolution (see Section \ref{sec3.3.2}), such 
uncertainty would be taken into consideration for MIRONG starting from normal galaxies without a reliable broad \ha\ detection in 
SDSS and only when the variation amplitude exceeds 0.17 dex do we think there is an upward or downward trend.

\begin{deluxetable}{cccccc}\label{variation-t}
\setlength{\tabcolsep}{0.06in}
\tablecaption{\it Summary of Emission-line Variability}
\tablewidth{0pt}
\tablehead{
\colhead{Class} & Number & \colhead{\hab} & \colhead{\hab-CL} & None & Percentage  \\
\colhead{(1)} & \colhead{(2)} & \colhead{(3)}  & \colhead{(4)}  & \colhead{(5)}  & \colhead{(6)} }
\startdata
Starforming & 15 & 4  &  1  & 10   & 33.3\% \\
Composite   & 20 & 3  &  4  & 12  & 35.0\% \\
Seyfert~1   &  4 & 2  &  2  & 0   & 100\%  \\
Seyfert~2   &  5 & 2  &  1  & 2   & 60.0\%   \\
LINER       &  9 & 2  &  1  & 6   & 33.3\% \\
Total       & 53 & 13 &  9 & 31  & 41.5\%  
\enddata
\tablecomments{
The spectroscopically observed sample has been classified into three types 
(columns 3-5) based on the variability of broad \ha\ and coronal lines,
noting that there is none with only coronal line variability.
Column~(1): BPT class from SDSS spectra, which was collected from Paper I.
Column~(2): the number of objects in each class.
Column~(3): the number of objects with variations in broad \ha\ but not coronal lines.
Column~(4): the number of objects with variations in both broad \ha\ and coronal lines.
Column~(5): the number of objects with neither broad \ha\ nor coronal line variations.
Column~(6): the percentage of sources that show emission-line variability 
(\hab\ or \hab-CL) in the given class.
}
\end{deluxetable}

\subsubsection{The Variability in Different Galaxy Types} 

First, the percentage of objects that display emission-line variability differs greatly in different spectral types (see Table~\ref{variation-t}). 
It is remarkable yet not surprising to note that Seyfert galaxies possess the highest
fraction given that AGNs display ubiquitous variability in both continuum and emission lines. In particular, all four Seyfert~1 galaxies 
show obvious \hab\ variations, with coronal lines being detected in two of them. Though obvious variability is not expected for normal 
Seyfert~2 galaxies (e.g.,\Citealt{Yip2009}), the line variability found for the Seyfert~2 galaxies in our sample is understandable, 
because there is a high possibility for them to be changing-look AGNs considering the selection criteria. In contrast to Seyfert galaxies, 
the star-forming (SF) and composite galaxies present significantly lower portion of variability (33.3\% and 35\%, respectively), with the former 
dominated by \hab\ while the latter is dominated by \hab-CL variations. 
Low-ionization nuclear emission-line regions (LINERs), which are usually considered to be photoionized by a weak AGN 
(\citealt{LINER1,LINER2}), hold the ratio same as star-forming galaxies though.

\begin{deluxetable*}{cccccccc}
\tablewidth{\textwidth}
\tablecaption{\it Summary of sources with emission-line variations}
\label{Summary-t}
\tablehead{
\colhead{Name} & \colhead{BPT} &\colhead{Time Coverage}&\colhead{$\rm H_{\alpha,B}$ Behavior} & \colhead{Iron CLs} & \colhead{$\rm He\,\textsc{II}\lambda4686$} & \colhead{$\rm N\,\textsc{III}\lambda4640$}&\colhead{Interpretation}  \\
\colhead{(1)} & \colhead{(2)} &\colhead{(3)}  & \colhead{(4)}  & \colhead{(5)}  & \colhead{(6)} & \colhead{(7)} &\colhead{(8)}}
\startdata
J0205+0004  &   SF  &       2017-2021 & Restored    & $\times$ & $\times$   & $\times$      & TDE	\\
J0859+0922  &	SF	&	2018-2021 & Declining	& $\times$ & $\times$	& $\times$	& TDE	\\
J1549+3327  &	SF	&	2017-2020 & Restored	& $\surd$  & $\surd$	& $\times$	& TDE	\\
J1620+2407  &	SF	&	2019-2021 & Declining	& $\times$ & $\times$	& $\times$	& TDE	\\
J1647+3843  &	SF	&	2018-2021 & Declining	& $\times$ & $\times$	& $\times$	& TDE	\\
J1043+2716  &	Composite &	2017-2021 & Declining	& $\surd$  & $\times$	& $\times$	& TDE  \\
J1111+5923  &   Composite & 2018-2020 & Declining   & $\times$ & $\times$  & $\times$ & TDE \\
J1442+5558  &	Composite &	2017-2021 & Maintain	& $\surd$  & $\surd$	& $\surd$	& Turn-on  \\
J1513+3111  &	Composite &	2017-2021 & Declining	& $\surd$  & $\surd$	& $\surd$   	& TDE	\\
J2203+1124  &	Composite &	2017-2019 & Declining	& $\surd$  & $\surd$	& $\times$	& TDE	\\
J1315+0727  &   Composite(b) & 2017-2017 & Rising     & $\times$ & $\times$    & $\times$      & AGN Flare\tablenotemark{\footnotesize \rm 3} \\
J1332+2036  &	Composite(b) &	2017-2021 & Declining	& $\times$ & $\times$	& $\times$	& AGN Flare \\
J1133+6701  &	LINER(b)    &	2018-2021 & Restored	& $\times$  & $\times$	& $\times$	& AGN Flare \\
J1115+0544  &	LINER	&	2016-2021 & Restored	& $\surd$  & $\times$	& $\times$	& TDE	\\
J1632+4416  &	LINER	&	2017-2018 & Restored	& $\times$ & $\times$	& $\times$	& TDE	\\
J1003+0202  &	Seyfert 2 &	2018-2021 & Maintain	& $\times$ & $\times$	& $\times$	& Turn-on	\\
J1238+0815  &	Seyfert 2&	2017-2017 & Declining	& $\times$ & $\times$	& $\times$	& AGN Flare	\\
J1657+2345  &	Seyfert 2&	2017-2021 & Declining	& $\surd$  & $\surd$	& $\times$	& TDE	\\
J0120-0829  &	Seyfert 1(b) &	2017-2018 & Restored\tablenotemark{\footnotesize \rm 2} & $\times$	& $\times$ & $\times$ & AGN Flare  \\
J1105+5941  &	Seyfert 1(b) &	2017-2021 & Declining 	& $\surd$  & $\surd$	&   $\times$	&  TDE \\
J1402+3922  &	Seyfert 1(b) &	2017-2021 & Restored\tablenotemark{\footnotesize \rm 1} & $\surd$	& $\surd$  & $\times$ & TDE \\
J1537+5814  &	Seyfert 1(b) &	2017-2021 & Restored\tablenotemark{\footnotesize \rm 1} & $\times$	& $\times$ & $\times$ & AGN Flare \\
\enddata
\tablecomments{This is a summary table of the 22 sources with emission-line variation. 
Column (2): BPT class from SDSS spectra. The galaxy types followed by a ``b'' in parentheses indicate that apparent broad Balmer lines were detected in their SDSS spectra according to the criteria in Section \ref{sec3.2}. 
Column (3): the first and last years of the follow-up observation. 
Column (4): all of the 22 sources show enhancement or emergence of broad \ha, but their subsequent $\rm H_{\alpha,B}$ evolution behaviors are somewhat different, which could be generally divided into the following four types.(i)``Rising'': the \hab\ rises up continuously to the end of observations. (ii) ``Declining'': the \hab\ apparently fades after enhancement but it is still at a level higher than in the SDSS spectra in the latest observation. (iii) ``Restored'': the \hab\ shows a declining trend after the enhancement and it has almost returned back to the same level as SDSS now; (iv) ``Maintain'', the \hab\ stays at a high level without a notable declining trend after the enhancement. 
Column (5):``$\surd$'' means iron coronal lines were detected in the DBSP spectra while ``$\times$'' means not. 
Column (6):same as column (5) but gives the detection of $\rm He\,\textsc{ii}\lambda4686$.
Column (7):detection of $\rm N\,\textsc{iii}\lambda4640$ or not. 
Column (8):preliminary interpretation of the MIRONG based on the spectral evolution. 
}
\tablenotetext{1}{The \hab\ in the two Seyfert I galaxies returned to the pre-outburst level about 1 yr after the first spectroscopic follow-up, continued to fade out and almost disappeared completely in the last epoch.}
\tablenotetext{2}{The \hab\ had already declined to a level lower than in the SDSS spectrum after 5 months since brightening at the first epoch, yet it shows a rising trend again later, that is consistent with rebrightening in the MIR.}
\tablenotetext{3}{``AGN flare'' means we stopped at the result that these MIRONG were caused by AGN activity, without further preference for detailed physical processes behind (see \ref{sec4.3})}
\end{deluxetable*}

\begin{figure*}
\figurenum{3}
\centering
\begin{minipage}{0.9\textwidth}
\centering{\includegraphics[angle=0,width=1.0\textwidth]{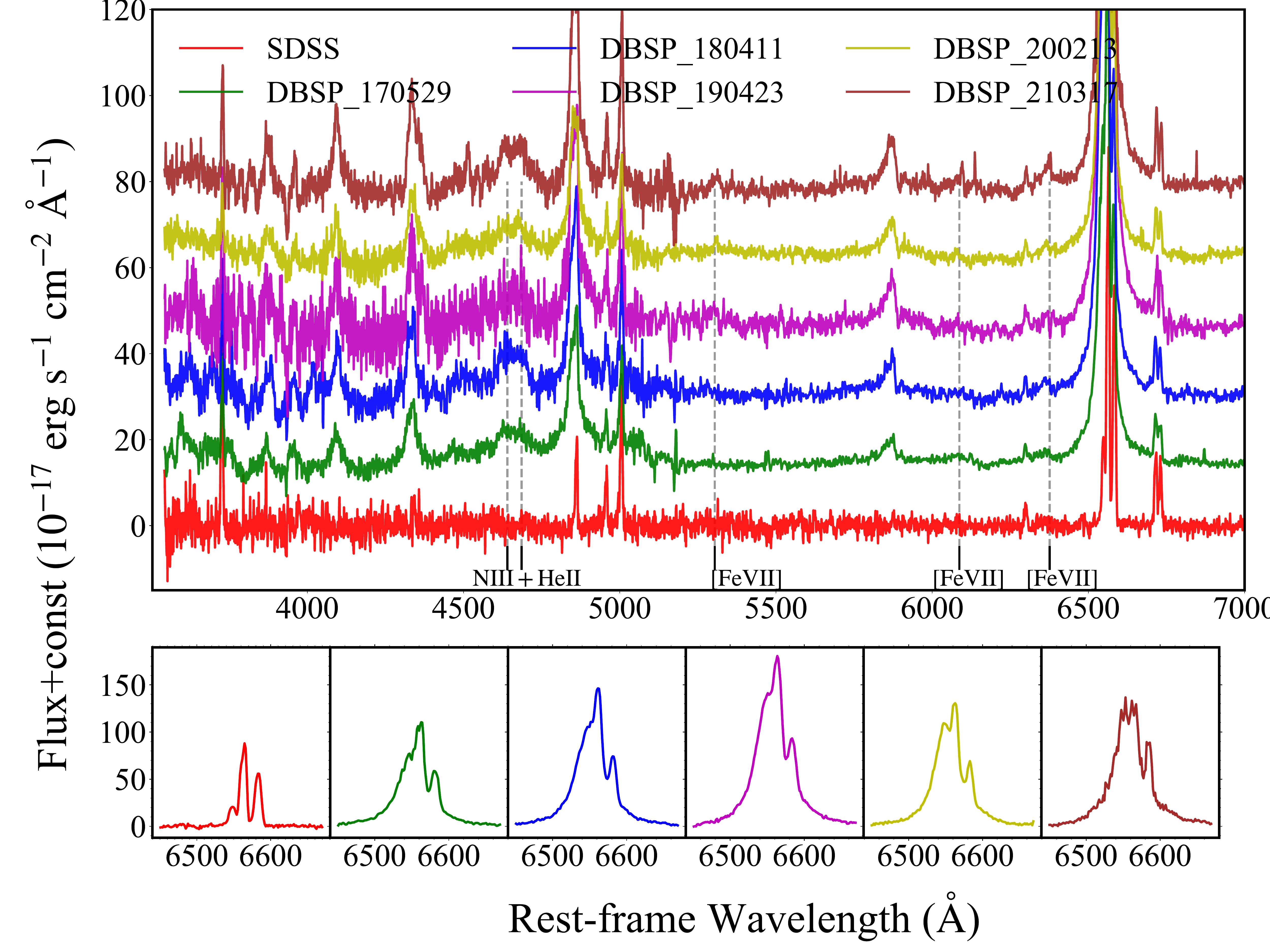}}
\end{minipage}
\caption{We show the spectral evolution of J1442+5558, including the whole spectral evolution (top panel, continuum subtracted) and the evolution of the \ha\ region (bottom panel) 
\label{J1442+5558_spec}}
\end{figure*}

\subsubsection{Long-term Evolution of Broad \ha\ and Coronal Lines}
\label{sec3.3.2}

All 22 objects with detection of emission-line variability with respect 
to SDSS spectra show an apparent increase in broad \ha\ emission while only 
a small fraction of them show coronal-line variations. Except for J1442+5558 and J1003+0202, 
the strength of \hab\ declines again in the subsequent observations 
although most of them have not returned back to the flux level of SDSS as of 
the latest epoch. It indicates that the \hab\ emission fades slowly and a minority of them
may sustain for a long time. Despite of their weakness, coronal lines clearly 
fade out after the first follow-up observation except for J1442+5558, whose 
coronal lines have just emerged in recent DBSP spectra while \hab\ has been 
detected in the first DBSP. We show the spectral evolution of J1442+5558 
in Figure~\ref{J1442+5558_spec}, and all 22 sources with emission line variation in Appendix~\ref{appendixB}.

\subsubsection{$\rm{N\,\textsc{iii}}$ and $\rm{He\,\textsc{ii}\lambda 4686}$}

During our spectroscopic analysis, we have noticed that some of them also 
show variations of $\rm He\,\textsc{ii}\lambda4686$ and Bowen line $\rm N\,\textsc{iii}\lambda4640$  
apart from broad Balmer lines and iron coronal lines. Particularly, 7 of the 10 \hab-CL sources present  $\rm He\,\textsc{ii}\lambda4686$, with two of them (J1442+5558 and J1513+3111) even with $\rm N\,\textsc{iii}\lambda4640$ emission. Both $\rm He\,\textsc{ii}\lambda4686$ and $\rm N\,\textsc{iii}\lambda4640$ lines are frequently detected in TDEs (\citealt{vV2021}).

\subsection{Sources without a Visible Spectral Change}
\label{sec3.4}

Among the 53 useful sources, 22 show 
obvious emission variability in at least broad \ha, while the other 31
appear unchanged in optical spectra according to our observations. 
It is necessary to explore whether the nondetection is intrinsic or simply
caused by selection effects. 
First, we have examined the S/N of the spectroscopic data of the variable
and nonvariable subsamples as the higher-quality data are undoubtedly more 
helpful for the detection of weak emission lines embedded in the continuum. 
However, no obvious difference is found.

As we have shown in section \ref{sec3.3.2}, the \hab\ and iron coronal lines 
of most objects show apparent decreasing trends after brightening.
Some of them have even disappeared rapidly, demanding quick spectral follow-ups to capture the variations. 
For this reason, if the spectroscopic observations were conducted too late, 
we might miss the fleeting emissions . 
We have calculated the time separation between the spectroscopic 
observational date and the first MIR brightening epoch 
($3\sigma$ brightening in either the $W1$ or $W2$ band; see details in Paper I).
There is no significant difference between the spectral variable and 
nonvariable subclass according to a Kolmogorov–Smirnov (K-S) test. 
It seems that the nondetection of spectral variability should not be a 
consequence of too-late observations if the two subclasses share comparable
decay timescales. 

\begin{figure*}[ht]
\figurenum{4}
\centering
\begin{minipage}{1\textwidth}
\centering{\includegraphics[angle=0,width=1.0\textwidth]{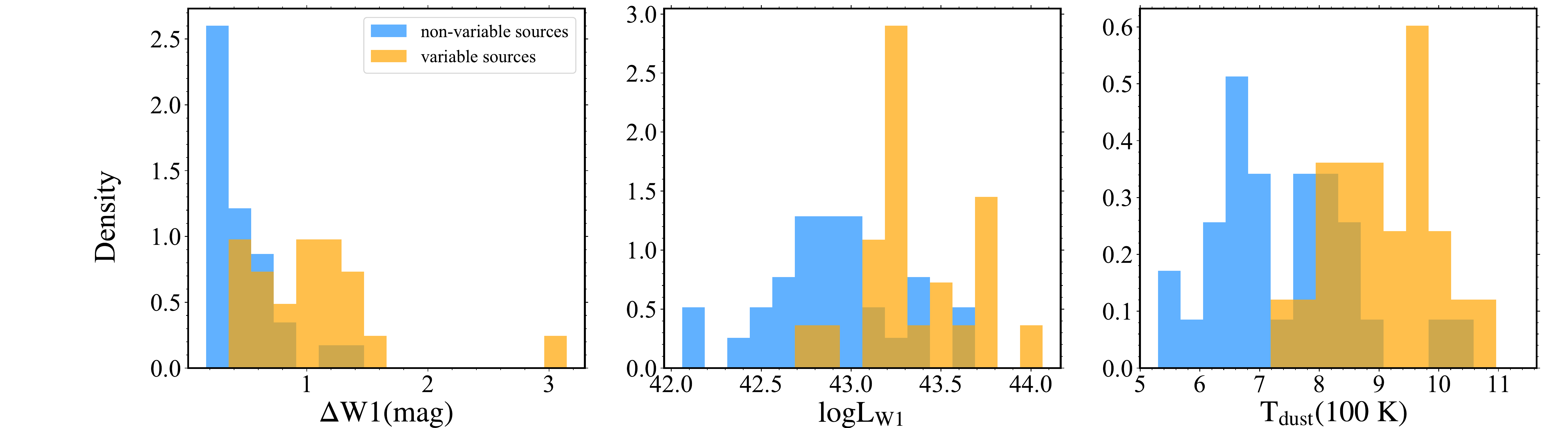}}
\end{minipage}
\caption{We show the comparison of the $W1$-band variability amplitude $\Delta W1$ (left panel), logarithmic $W1$ peak luminosity $logL_{W1}$ (middle panel), and dust temperature $T_{dust}$ (right panel). All data were drawn from Paper I, in which the dust temperature was picked out from results fitting with dust absorption coefficient.
\label{fig:comparison}}
\end{figure*}

Dust is ubiquitous in galaxies from galactic disks to galactic centers.
It is especially considerable for our MIRONG sample because they were selected 
by IR echoes radiated by the circumnuclear dust around SMBHs.
The dust along the light of sight can obscure the optical photons,
making them undetectable in both continuum and emission lines.
Figure~\ref{fig:comparison} shows the $W1$ band variability amplitude $\Delta W1$, 
$W1$ peak luminosity $L_{W1}$, dust temperature $T_{dust}$ 
(data drawn from Paper I, also a significant difference in $\Delta W2$ and $L_{W2}$ ). 
All three parameters show a significant difference
between the variable and nonvariable subclass based on a K-S test with the $p-$value being $3.1\times10^{-5},1.1\times10^{-5}$, and $3.8\times10^{-6}$, respectively.
In detail, the nonvariable sources have lower $\Delta W1$, $L_{W1}$ 
and $T_{dust}$, which can be explained by the self-obscuration of dust emission. 
We have checked the narrow-line Balmer decrements of the two populations and 
found no difference, indicating a similarity in galactic-scale dust. 
The dust discrepancy between them might be in the galactic nucleus. 

We have also looked into the distributions of other parameters, 
including \mbh\ and the rising and decay timescales of MIR light curves.
All show negligible differences.

\section{Discussion}

Optical spectra usually carry vital information for deciphering the nature 
of unknown transients. Aiming to explore the nature of the MIRONG sample,
we have conducted extensive follow-up observations. Prominent emission-line 
variability has been detected in 22 out of the 53 objects (see details in Table \ref{Summary-t}). 
By means of the spectral results, we will show that the SN scenario is 
further proved to be the less likely case, corroborating the associations
with SMBHs as proposed in paper I. 
Specifically, we will try to understand the MIR outburst under the context
of diverse transient accretion events of SMBHs, such as TDEs, turn-on AGNs, or 
sporadic gas accretion. Even for those without a visible spectral change,
some clues can also be acquired.

\subsection{SN Scenario Is Further Disfavored}

In fact, the SN scenario has been largely excluded 
(paper I) based on MIRONG's high MIR luminosity (see also \citealt{Wang2018}) 
and their proximity to the galaxy center. 
Spectroscopic evolution further supports this conclusion. 

First, the varying components in the optical spectra overwhelmingly 
feature emission but without notable absorption lines that are atypical
for conventional SNe (\citealt{Filippenko1997}). Further, the broad \ha\ lasts 
typically 2-4 yr or more from their first detection in our sample, which is also 
in sharp contrast with the fast decline for a typical Type II SN (e.g., \citealt{Ha_decline,SN2004et}). 
Only a subclass of SNe (Type IIn), which shows the signatures of the interaction
between the SN ejecta and the circumstellar medium and exhibits obvious 
Balmer emissions, most notably \ha, with a weak or absent broad absorption component (\citealt{Schlegel1990,Filippenko1997}), bears some 
similarities with our sample. And some objects ,such as  SN2010jl 
(\citealt{SN2010jl,SN2014ab_SN2010jl}), SN2015da (\citealt{SN2015da}), SN2005ip (\citealt{SN2006jd&SN2005ip,SN2005ip}), and SN2006jd (\citealt{SN2006jd&SN2005ip}), 
displayed a long-lasting broad \ha\ with similar line luminosity to our sample and could be accompanied 
by coronal lines. However, their late-time spectra usually show prominent low-ionization emission 
lines such as near-infrared $\rm Ca\,\textsc{ii}$ triplets and $\rm O\,\textsc{i}\lambda8446$, which are completely absent in our spectra. 
Hence, our spectroscopic results argue against an SN origin.

\subsection{MIRONG Occurring in Normal or Weak-AGN Galaxies} 

The final sample of the 22 spectroscopically variable sources consists of 
7 Seyfert galaxies according to their SDSS BPT classification. 
In addition, we note that there are three more galaxies in other classes, 
that is J1315+0727, J1332+2036 (composite), and J1133+6701 (LINER), that 
show evident broad \ha\ in their SDSS spectrum. 
Therefore, we grouped the 10 sources (Seyferts or with a broad-line)
into a subsample of unambiguous AGNs before the MIR outburst 
and the remaining 12 into a subsample without or with very weak AGNs.
We begin our discussion with the latter subsample.

\subsubsection{One Bona Fide Turn-on AGN Candidate}

The sudden and drastic brightening in the centers of normal galaxies is usually 
attributed to TDEs or turn-on AGNs once the SN scenario is ruled out. 
Although in the literature the term `turn-on AGN' was also used by some authors to refer to 
changing-look AGN which transits from Seyfert 2 or 1.8 to type 1.5 or 1 
(e.g., \citealt{MacLeod2016,Yang2018,WangJ2018}), we will strictly use the term 
to mean the transition from normal galaxies to Seyfert 1 galaxies or quasars. 
\citet{Gezari2017} reported the first case of a turn-on quasar, 
which transits from an absorption-line galaxy to a quasar in a surprisingly 
short time scale of a few months. 
\citet{Yan2019} presented a case of a transit from a LINER to a Seyfert 1 galaxy 
within a few years in SDSS 1115 discovered in the spectroscopic 
follow-up of the MIRONG sample. Both of these objects showed a plateau 
in the optical light curve following an initial brightening. 
Note that both objects displayed also strong IR echoes and are included in 
the MIRONG sample. Subsequently, six more transients from LINERs to 
broad-lined AGN-like spectra were discovered by ZTF \citep{Frederick2019}. 
Note that three of them also displayed weak broad emission lines 
in their pre-outburst spectra, which is similar to J1133+6701 in our sample. 
Nevertheless, most sources fade out in about a year, but with 
some fluctuations \citep[see Figure 5]{Frederick2019}, 
different from those of known optical TDEs. 
Together with large black hole masses, which are derived from the correlations
between \mbh\ and galactic properties, they appear more likely to be associated with 
nuclear activity than with TDEs, although the nature of these sources remains to be unraveled. 

However, as we have mentioned above, the appearing broad \ha\ in most objects 
fades immediately in the subsequent observations, indicating a short-lived accretion. 
Only J1442+5558 enters a long plateau phase in the light curve of \hab, 
which has lasted for at least four years, and thus is a good candidate of a bona fide turn-on AGN. 
However, the integrated radiative energy to now for J1442+5558 is $1.1\times10^{52}$~erg, 
well within the range expected for TDEs.  
Thus, further monitoring was also necessary to determine whether the \hab\ continues 
to stay at the high state or is going to decrease soon.
In fact, J1115+0544, another object in our sample, has been thought of as a 
reliable bona fide turn-on AGN  because its \hab\ has steadily been 
there for at least two years (2016-2018; \citealt{Yan2019}).
However, the latest observation performed at UT 2021-01-07 shows that its
\hab\ has almost disappeared, to our surprise. 
%
\subsubsection{Most are Good TDE Candidates}

Therefore, the bona fide turn-on AGNs should be minorities in our sample 
since only one shows persistent \hab\ emission after the appearance.
The remaining 11 are formally consistent with TDE-like transients. 
In particular, the \hab\ of J0205+0004, J1549+3327, and J1632+2345 has 
vanished rapidly within one year and  J1115+0544 has also returned back to 
the flux level of SDSS though the strong broad \ha\ has remained for at least 2 yr. 
Further monitoring of the other seven objects in the declining
phase will inform us whether their \hab\ would also fade away completely.
We need to emphasize that the behavior of the \hab\ evolution is certainly not 
sufficient for concluding the nature of MIRONG although it can reflect the 
energy and timescale of the primary emission to some extent. 
Actually, our knowledge of the spectral characteristics accompanied by 
various SMBH transients is still in the accumulation stage and awaiting 
in-depth understanding. 

Besides the above-mentioned \hab\ emission, it is notable that 
iron coronal lines have been detected in 5 of the 11 sources, 
with 3 also being detected with $\rm He\,\textsc{ii}\lambda4686$
and 1 being detected with $\rm N\,\textsc{iii}\lambda4640$.
All of these emission lines are previously found and used to claim a TDE.
Remarkably, the ZTF group has recently classified their uniformly selected 
TDE sample into three categories based on whether or not 
the hydrogen Balmer lines,  $\rm He\,\textsc{ii}\lambda4686$ and $\rm N\,\textsc{iii}\lambda4640$ are present (\citealt{vV2021}), 
although the physics that drives the diversity is still poorly understood.
Aside from the TDEs found by optical photometric surveys, 
the spectroscopic selection by extreme iron coronal lines in normal galaxies 
has also been proven to be a valid approach (\citealt{Komossa2008,Wang2011,Wang2012}), theoretically supported by the anticipation that TDEs can nicely provide the
soft X-ray-ionizing photons. Their transient nature has been further confirmed
by the long-term spectral evolution (\citealt{Yang2013}), dust IR echoes 
(\citealt{Dou2016}), and soft X-ray radiation (\citealt{He2021}).
In terms of the similarity of prominent IR echoes and high detection rate of
coronal lines, our sample could be most likely to be analogous to the 
coronal-line selected TDEs.

In brief, the appearance and fading of \hab, iron coronal lines, 
$\rm He\,\textsc{ii}\lambda4686$ and $\rm N\,\textsc{iii}\lambda4640$ are all nice evidence 
for TDEs although the real benchmark of the TDE spectrum is still poorly known.

\subsubsection{The Possibility of Sporadic Gas Accretion}

Our MIRONG sample is very different from optical TDEs in IR echoes. A systematic investigation of the MIR light curves of optical TDEs 
suggest that their echoes are very weak or negligible, yielding a very low dust-covering factor ($f_c$) of those SMBHs (\citealt{Jiang2021b,vV2016,vV2021b}). 
On the contrary, the high MIR luminosity of MIRONG definitely gives a much higher $f_c$  (\citealt{Jiang2021a}). The locally gas-rich 
environment means that the SMBH will not likely starve due to lack of gas supply permanently. Is it possible that the sporadic gas 
accretion or instability in the accretion flow triggers the transient accretion? 

First, the instability of the gas reservoir in the outer disk may result in sporadic feeding to the central black hole. Such a process 
is invoked to explain changing-look AGNs. Radiation pressure instability in the inner accretion disk with a very low accretion rate is 
exceedingly longer than the observed timescale \citep{Gezari2017}.  While standard limit-cycle instability, as in cataclysmic variables, 
in the outer disk can be ruled out because of its very long timescale \citep{Lin1986}, a modified version of the limit-cycle instability 
model was proposed for changing-look AGNs \citep{Sniegowska2020}. In the quiescent state, the accretion rate is low, and the system 
consists of a truncated outer thin disk and an inner hot advection-dominated flow. The instability takes place in a narrow ring of the 
thin disk on the boundary of two accretion flows at a few tens of the Schwarzschild radius. Although the model is still under development, 
initial analysis of the stability of a narrow ring does suggest it is able to reproduce the short transit timescale of months to years 
observed in changing-look AGN as well as in our objects. The duty cycle increases with increasing steady accretion rate, viscosity, and 
width of the ring. A low steady accretion rate, and thus low duty cycle, may explain the lack of a prominent AGN signature in the 
narrow-line spectra, although further detailed modeling is required to confirm this.   

Second, stochastic feeding by a cold gas cloud, such as G1/2 clouds in the Galactic center, may eventually lead to an outburst. The 
presence of thick dust components on subparsec to parsec scales as revealed by IR echoes indicates the presence of a rich cold gas 
environment, analogously with a circumnuclear molecular disk at the Galactic center, in which the G2 cloud originated \citep{Schartmann2015}. 
Although the encounter of G2 with the supermassive black hole did not significantly increase the accretion rate from both observations 
and numerical simulations \citep{Morsony2017}, strong mass feeding may be expected if a dense thin disk is present. In that case, the 
collision between the cloud and dense gas in the disk may result in significant dissipation, likely ending in gas accretion. For a 
high plunged cloud, tidal disruption will further increase the cross section of collision, enhancing such fueling process. However, as 
the encounter of a cloud takes place at $10^2-10^3 r_s$, and the accretion time is expected long if the disk remains cold. Nevertheless, 
the encounters with clouds are random. If the directions of angular momentum of the disk and the cloud are opposite, the collision may 
leave a significant fraction of very low-angular-momentum gas, which can be quickly accreted.  

\subsection{MIRONG Occurring in AGNs}
\label{sec4.3}

The real physics behind MIRONG in active galaxies could be complicated and 
diverse, with turn-on AGNs, unusual AGN variability, and TDEs all possible. 
Distinguishing between them is quite challenging to our current knowledge.
We will try to give our initial impression of them here, 
leaving a detailed analysis to our subsequent works.

The amplitude of stochastic variability, which is a defining characteristic of AGNs, 
is generally not higher than tenths of a magnitude within months to years although it 
increases toward longer timescales (e.g.,\citealt{VandenBerk2004,MacLeod2010}). 
Only the most extreme tail of AGN variability distribution may show an
amplitude comparable with TDEs over a short period of time 
(e.g., \citealt{MacLeod2012,Rumbaugh2018}).
Moreover, conventional AGN variability does not show regular and smooth
patterns, such as the power-law decay seen in TDEs. 
Nevertheless, there indeed exists a population of AGNs displaying major flares
at timescales of years, which resembles TDEs more, in their light curves,
(e.g., \citealt{Graham2017,Kankare2017,Trakhtenbrot2019}).
The nature of these AGN flares remains very elusive even with spectroscopic
observations as we still lack adequate information to distinguish each 
scheme from impostors at the benchmark level (\citealt{Zabludoff2021}), 
as we emphasized in the last section.

Exploration of our MIRONG sample becomes even difficult without 
direct variability information of their original emission. 
Luckily, emission-line variability, which can be considered as an alternative 
tracer, has been detected in most objects (see Table~\ref{variation-t}). 
The detection of \hab\ variation itself is not a surprise for AGNs, 
yet their rapid evolution may shed light on the central ionizing sources. 
Characteristic emissions other than \hab, namely iron coronal lines
and $\rm He\,\textsc{ii}\lambda4686$, might give further clues to the outburst.
Only three objects (J1105+5941, J1402+3922, and J1657+2345) show these
emissions, and their \hab\ has either been restored or is steadily declining.
We thus take the three as tentative TDEs in AGNs for the time being
as both iron coronal and $\rm He\,\textsc{ii}\lambda4686$ are representative features of
TDE spectra, although not decisive. 

For the remaining sources, only \hab\ variability has been found and their
physics are even harder to constrain. To be consistent with prior definitions, 
we categorize J1003+0202, which was a Seyfert 2 and presented stable \hab\ 
after the brightening, as a turn-on subclass while the rest sources as AGN flares.  
We note that J1315+0727 only has two observations in 2017 and its \hab\ was
still rising. More recent spectra will confirm whether it has faded away now.
It should be emphasized again that the flares here may be caused by 
distinct processes in different sources, such as TDE, sporadic gas accretion, 
or special accretion disk instability.

\subsection{MIRONG without Spectral Variation}

The nature of the other 31 objects without clear spectral variations 
is still mysterious, yet might be even more fascinating.  
One possibility is that they belong to a new population 
of fast transients, which evolve much faster than variable sources. 
The high-cadence surveys over the past decade have indeed revealed a population
of transients with timescales obviously shorter than those of SNe and TDEs, and 
are simply called "Rapidly Evolving Transients" or "Fast-Blue Optical Transients" (FBOTs)
in literature (e.g., \citealt{Drout2014,Pursiainen2018}) as their physical
origin remains a puzzle. 
For instance, the nearest and most well-studied case, AT~2018cow, shows a
rise-to-peak timescale of only three days and faded away within a couple of months
(\citealt{Margutti2019,Perley2019}). 
The reprocessed MIR light curves cannot reflect the intrinsic outburst timescales
unambiguously due to their coupling with dust distributions. 
Moreover, the optical surveys that overlapped with the outbursting period of our MIRONG
sample have either poor cadence (e.g.,  PanSTARRS, PTF) or depth (e.g., ASASSN), 
thus they cannot put effective constraints on their optical emission.
For likely the same reason, most of them have been missed by optical surveys. 
Future selection of more recent MIRONG, such as the outbursts after 2018
with both excellent WISE and ZTF light curves, may help test the 
fast transient hypothesis as the ZTF survey has a much higher cadence and depth 
compared with previous surveys. 

The discrepancy in the dust properties (see Section \ref{sec3.4}) suggests that the 
dust obscuration may play a vital role and those spectral
nonvariable sources correspond to more seriously obscured ones.
In this scenario, they are good candidates of obscured TDEs, turn-on AGNs, 
or any other types of SMBH transients that are undetectable in optical bands.
The discrepancy in IR emission can be also partly accounted for by the possibility that 
the outburst energy of the nonvariable sources is systematically 
lower than that of variable sources if their dust contents (covering factor) are identical. 
Further investigations with the aid of multiwavelength observations (e.g., radio;
\citealt{Mattila2018}) are needed to make a conclusion about whether they
are real dust-obscured events or an intrinsically distinct population.

\section{Conclusions}

As the first and essential step to identifying the nature of MIRONG, which are selected by a blind 
search, spectroscopic observations yield much valuable information
of the outbursts. The main results and conclusions are summarized as below.

\begin{enumerate}
    \item[1.] We have obtained optical spectra for 54 objects (53 useful) in the MIRONG sample since 2017. Among them, 22 show the appearance or significant brightening of the broad \ha\ (\hab) emission, corroborating the primary conclusion drawn from Paper I, that is the MIR outbursts are mainly connected with dust echoes of transient SMBH accretion events. Regarding the difference types of galaxies, Seyferts have the highest ratio of \hab\ variability while a third of the other types (SF, composite, and LINER) also show an obvious change in \hab. 
    
    \item[2.] Multiepoch spectroscopic observations have been continuously carried out for targets with the detection of \hab\ variability until the broad \ha\ component has disappeared or been restored. Most of them display a declining trend immediately after the brightening, suggesting short-lived transients, such as TDEs, sporadic gas accretion, or special AGN variability. Only two show long-term stable \hab\ emission, which is more in agreement with the turn-on AGN scenario. Particularly, the composite galaxy J1442+5558 is an excellent candidate to be a bona fide turn-on AGN, which is hitherto extremely rarely found.
  
    \item[3.] The characteristic iron coronal lines, $\rm He\,\textsc{ii}\lambda4686$ and $\rm N\,\textsc{iii}\lambda4640$ have also been detected in quite a fraction of these objects, which could provide further clues to their nature. We have naively classified the sources with multiple transient emissions as TDE candidates, particularly for those occurring in inactive galaxies. The real origin of outbursts in AGNs could be diverse and challenging to distinguish from each other as they still lack an accepted benchmark in spectra to our knowledge.
    
    \item[4.] Sources without spectral variations do not behave differently in most distributions. The lower MIR luminosity and dust temperature of the nonvariable sources could be a result of self-obscuration although intrinsic weakness is also possible. If those objects are indeed more obscured TDEs, turn-on AGNs, or any other types of SMBH transients, they are worthy of further studies with the aid of observations from other bands (e.g., radio). Another explanation could be that they belong to a new population of fast transients yet without compelling evidence currently.
    
\end{enumerate}

The nature of MIRONG has been further joined with SMBH transient accretion events based on spectral results while their specific physics remains ambiguous. Actually, the community is still vague as to how to distinguish between different types of nuclear transients accurately. A more detailed analysis of individual sources or subsets  will be presented in our subsequent papers. Regardless of what is the true case eventually, MIRONG must be be an indispensable part of mounting SMBH transient events, i.e., as a substantial supplement for the optical sample. The NEOWISE survey is still in commission and should yield out more MIRONG after our primary selection (\citealt{Jiang2021a}). The more recent MIRONG sample nicely meets the golden era of time-domain surveys at multiple wavelengths, such as ZTF (\citealt{Bellm2019}), LSST (\citealt{Ivezic2019}), and WFST in the optical band and eROSITA (\citealt{Merloni2012}) and the Einstein Probe (\citealt{Yuan2015}) in the X-ray band. We have the opportunity to perform a more comprehensive study with the new MIRONG sample and to move forward in understanding the mysterious events conclusively.

\acknowledgments
We thank the referee for helpful comments and suggestions, which led to the improvement of the paper. This work is supported by the B-type Strategic Priority Program of the Chinese Academy of Sciences (Grant No. XDB41000000), 
Chinese Science Foundation (NSFC-11833007, 12073025, 11421303, 12103048, 11822301),China Manned Spaced Project (CMS-CSST-2021-B11) and the Fundamental Research Funds for the Central Universities.
This research uses data obtained through the Telescope Access Program (TAP).
Observations obtained with the Hale Telescope at Palomar Observatory 
were obtained as part of an agreement between the 
National Astronomical Observatories, Chinese Academy of Sciences, 
and the California Institute of Technology.

\clearpage
\onecolumngrid
\begin{appendices}
\section{ Summary of spectroscopic follow-up observations \label{appendixA} }
We show the spectroscopic follw-up osbervations and some basic information in the Table \ref{inf_table} at this appendix. 
\startlongtable
\begin{deluxetable*}{lllllll}
\tablewidth{\textwidth}
\tablecaption{Summary of spectroscopic follow-up observations\label{inf_table}}
\tablehead{name&redshift&Type&telescope&observation date&exposure time&slit width\\
    &        &         & & YYMMDD&s &arcsec}
\startdata
SDSSJ010320.42+140149.8&0.0418&StarForming&P200&171216&1200&1.5\\
SDSSJ012047.99-082918.4&0.0347&Seyfert 1(b)&P200&170630&600&1.5\\
&&&P200&171124&1500&1\\
&&&P200&180105&1200&1\\
SDSSJ020552.16+000411.8&0.0765&StarForming&P200&171111&1200&1.5\\
&&&P200&171124&600&1\\
&&&P200&210107&900&1.5\\
SDSSJ081403.78+261144.3&0.0757&StarForming&P200&171026&1200&1.5\\
&&&P200&171124&900s&1\\
SDSSJ083536.49+493542.7&0.0424&Composite&P200&200213&1200&1.5\\
SDSSJ085959.46+092225.6&0.1519&StarForming&P200&180224&1800&1.5\\
&&&P200&200213&1800&1.5\\
&&&P200&210317&1800&1\\
SDSSJ090924.55+192004.8&0.1072&Composite&P200&170420&900&1.5\\
&&&P200&170529&900&1.5\\
&&&P200&171124&900&1\\
SDSSJ094303.26+595809.3&0.0749&LINER&P200&180224&1200&1.5\\
SDSSJ094456.56+310552.2&0.0346&Composite&P200&200213&1200&1.5\\
SDSSJ100350.97+020227.6&0.1247&Seyfert 2&Lick/Shane&180318&1200&1.5\\
&&&P200&190423&1200&1.5\\
&&&P200&200213&1200&2\\
&&&P200&210317&1800&1\\
SDSSJ103753.68+391249.6&0.1068&StarForming&P200&200213&1800&1.5\\
SDSSJ104306.56+271602.1&0.1281&Composite&P200&170420&900&1.5\\
&&&P200&170525&1560&1.5\\
&&&P200&171111&2100&1.5\\
&&&P200&171124&900&1\\
&&&P200&180224&1800&1.5+2.0\\
&&&P200&180411&1500&1\\
&&&P200&210317&1800&1\\
SDSSJ105801.52+544437.0&0.1306&StarForming&P200&200213&1200&1.5\\
SDSSJ110501.98+594103.5&0.0337&Seyfert 1(b)&P200&170525&900&1.5\\
&&&P200&171111&1200&1.5\\
&&&P200&171124&800&1\\
&&&P200&180224&1200&1.5\\
&&&P200&180411&800,700&1\\
&&&P200&200213&1200&2\\
&&&P200&210317&1200&1\\
SDSSJ110958.34+370809.6&0.026&LINER&P200&170525&900&1.5\\
&&&P200&171124&600&1\\
&&&P200&210317&1200&1\\
SDSSJ111122.44+592334.3&0.1697&Composite&Lick/Shane&180318&1800&1.5\\
&&&P200&190423&1200&2\\
&&&P200&200213&1200&1.5\\
SDSSJ111536.57+054449.7&0.09&LINER&LAMOST&160106&5400&3\tablenotemark{\footnotesize a}\\
&&&P200&170525&1500&1.5\\
&&&P200&171111&1200&1.5\\
&&&P200&171124&1200&1\\
&&&P200&180114&1200&1.5\\
&&&P200&180408&1200&1.5\\
&&&P200&180411&1800&1\\
&&&Keck&180513&600&1\\
&&&P200&210107&1200&1.5\\
SDSSJ112018.31+193345.8&0.1279&Composite&P200&170525&1200&1.5\\
&&&P200&171124&900&1\\
SDSSJ112446.21+045525.4&0.074&Composite&P200&180224&1200&1\\
&&&P200&180408&1800&1.5\\
SDSSJ112916.12+513123.5&0.0329&Composite&P200&170529&900&1.5\\
&&&P200&171124&900&1\\
SDSSJ113355.93+670107.0&0.0397&LINER(b)&Lick/Shane&180511&1800&1.5\\
&&&P200&190423&1200&2\\
&&&P200&200213&1200&1.5\\
&&&P200&210317&1800&1\\
SDSSJ113901.27+613408.5&0.1346&StarForming&Lick/Shane&180318&1800&1.5\\
&&&P200&180408&1200&1.5\\
SDSSJ120338.31+585911.8&0.0469&StarForming&Lick/Shane&180612&1800&1\\
SDSSJ121907.89+051645.6&0.0825&Composite&Lick/Shane&180318&1200&1.5\\
SDSSJ123852.87+081512.0&0.1138&Seyfert 2&P200&170529&900&1.5\\
&&&P200&171124&600&1\\
SDSSJ124521.42-014735.4&0.2154&LINER&P200&180408&1200&1.5\\
SDSSJ130532.91+395337.9&0.0725&LINER&Lick/Shane&180318&1200&1.5\\
SDSSJ130815.57+042909.6&0.0483&Seyfert 2&P200&180224&600&1\\
SDSSJ131509.34+072737.6&0.0918&Composite(b)&P200&170529&900&1.5\\
&&&P200&171124&600&1\\
SDSSJ132259.94+330121.9&0.1269&LINER&P200&180408&1200&1.5\\
SDSSJ132902.05+234108.4&0.0717&Composite&Lick/Shane&180318&1800&1.5\\
&&&Lick/Shane&180613&1800&1.5\\
SDSSJ133212.62+203637.9&0.1125&Composite(b)&P200&170622&600&1.5\\
&&&P200&171124&600&1\\
&&&P200&180411&800,850&1\\
&&&P200&190423&1200&2\\
&&&P200&190621&1200&1.5\\
&&&P200&210317&1200&1\\
SDSSJ140221.26+392212.3&0.0637&Seyfert 1(b)&P200&170622&600&1.5\\
&&&P200&171124&600&1\\
&&&P200&180408&1200&1.5\\
&&&P200&180411&1200&1\\
&&&Lick/Shane&180612&1800&1\\
&&&P200&200213&1800&1.5\\
&&&P200&210107&1200&1.5\\
SDSSJ140648.43+062834.8&0.0845&StarForming&P200&170525&1200&1.5\\
&&&P200&190423&1200&2\\
SDSSJ141235.89+411458.5&0.1025&Composite&P200&180408&1200&1.5\\
SDSSJ144227.57+555846.3&0.0769&Composite&P200&170529&900&1.5\\
&&&P200&180224&1200&1\\
&&&P200&180411&1200,1220&1\\
&&&P200&180607&1200&1\\
&&&P200&190423&1200&2\\
&&&P200&190621&1800&1.5\\
&&&P200&200213&1200&1\\
&&&P200&210317&1200&1\\
SDSSJ150844.22+260249.1&0.0826&Composite&Lick/Shane&180613&1800&1.5\\
SDSSJ151257.18+280937.5&0.1155&LINER&P200&180224&600&1\\
&&&Lick/Shane&180613&1800&1.5\\
SDSSJ151345.76+311125.0&0.0718&Composite&P200&170602&1500&1.5\\
&&&P200&180411&1200,1230&1\\
&&&P200&180607&1200&1\\
&&&P200&190423&1200&2\\
&&&P200&210317&1200&1\\
SDSSJ153711.29+581420.2&0.0936&Seyfert 1(b)&P200&170622&600&1.5\\
&&&P200&180411&1200,1230&1\\
&&&Lick/Shane&180612&1800&1\\
&&&P200&190621&1600&1.5\\
&&&P200&200213&1800&1.5\\
&&&P200&210107&1500&1.5\\
SDSSJ154955.19+332752.0&0.0857&StarForming&P200&170602&900&1.5\\
&&&P200&180411&1200,1230&1\\
&&&P200&180607&1200&1\\
&&&P200&190423&1200&2\\
&&&P200&200213&1800&1.5\\
SDSSJ155437.26+525526.4&0.0664&Composite&Lick/Shane&180613&1800&1.5\\
SDSSJ155539.95+212005.7&0.0709&Composite&Lick/Shane&180318&1800&1.5\\
&&&Lick/Shane&180613&1800&1\\
SDSSJ155743.52+272753.0&0.0316&Composite&Lick/Shane&180318&1800&1.5\\
SDSSJ160052.26+461242.9&0.1974&StarForming&P200&190621&1200&1.5\\
SDSSJ162034.99+240726.5&0.0655&StarForming&P200&190621&1200&1.5\\
&&&P200&210317&900&1\\
SDSSJ162810.03+481047.7&0.1245&StarForming&Lick/Shane&180612&1800&1\\
SDSSJ163246.84+441618.5&0.0579&LINER&P200&170602&900&1.5\\
&&&Lick/Shane&180511&1800&1.5\\
SDSSJ164754.38+384342.0&0.0855&StarForming&Lick/Shane&180318&1800&1.5\\
&&&Lick/Shane&180612&1800&1\\
&&&P200&210317&1200&1\\
SDSSJ165726.81+234528.1&0.0591&Seyfert 2&P200&170602&900&1.5\\
&&&P200&180607&1200&1.5\\
&&&P200&190423&1800&1.5\\
&&&P200&190621&1800&1.5\\
&&&P200&210317&1800&1\\
SDSSJ165922.65+204947.4&0.0451&Seyfert 2&Lick/Shane&180318&1200&1.5\\
&&&Lick/Shane&180511&1800&1.5\\
SDSSJ214603.88+104128.6&0.1636&StarForming&P200&190621&1200&1.5\\
SDSSJ220349.23+112433.0&0.1863&Composite&P200&170628&1200&1.5\\
&&&P200&171111&2400&1.5\\
&&&P200&171124&1500&1\\
&&&Lick/Shane&180612&1800&1\\
&&&P200&190621&1200&1.5\\
SDSSJ221541.60-010721.0&0.0477&Composite(b)&Lick/Shane&180612&1200&1\\
\hline
\enddata
\tablecomments{The host galaxy spectral type in the third column was collected from Paper I, and 
the galaxy types followed ``b'' in bracket indicated that apparent broad balmer lines were detected in their $\sdss$ according to the criteria in \ref{sec3.2}. At the most time, the exposure time displayed in the sixth column was the same with each other for red and blue end observation
,but sometimes their may be a little difference, for which in the table the left one represented the blue end exposure time}
\tablenotetext{a}{The $3^{\prime \prime}$ given here for the LAMOST spectrum is the spectral fiber diameter.}
\end{deluxetable*}

\clearpage
\onecolumngrid
\section{Spectral evolution of the 21 sources with emission lines variation \label{appendixB}}
\begin{figure*}
\figurenum{5}
\centering
\begin{minipage}{0.9\textwidth}
\centering{\includegraphics[angle=0,width=1.0\linewidth]{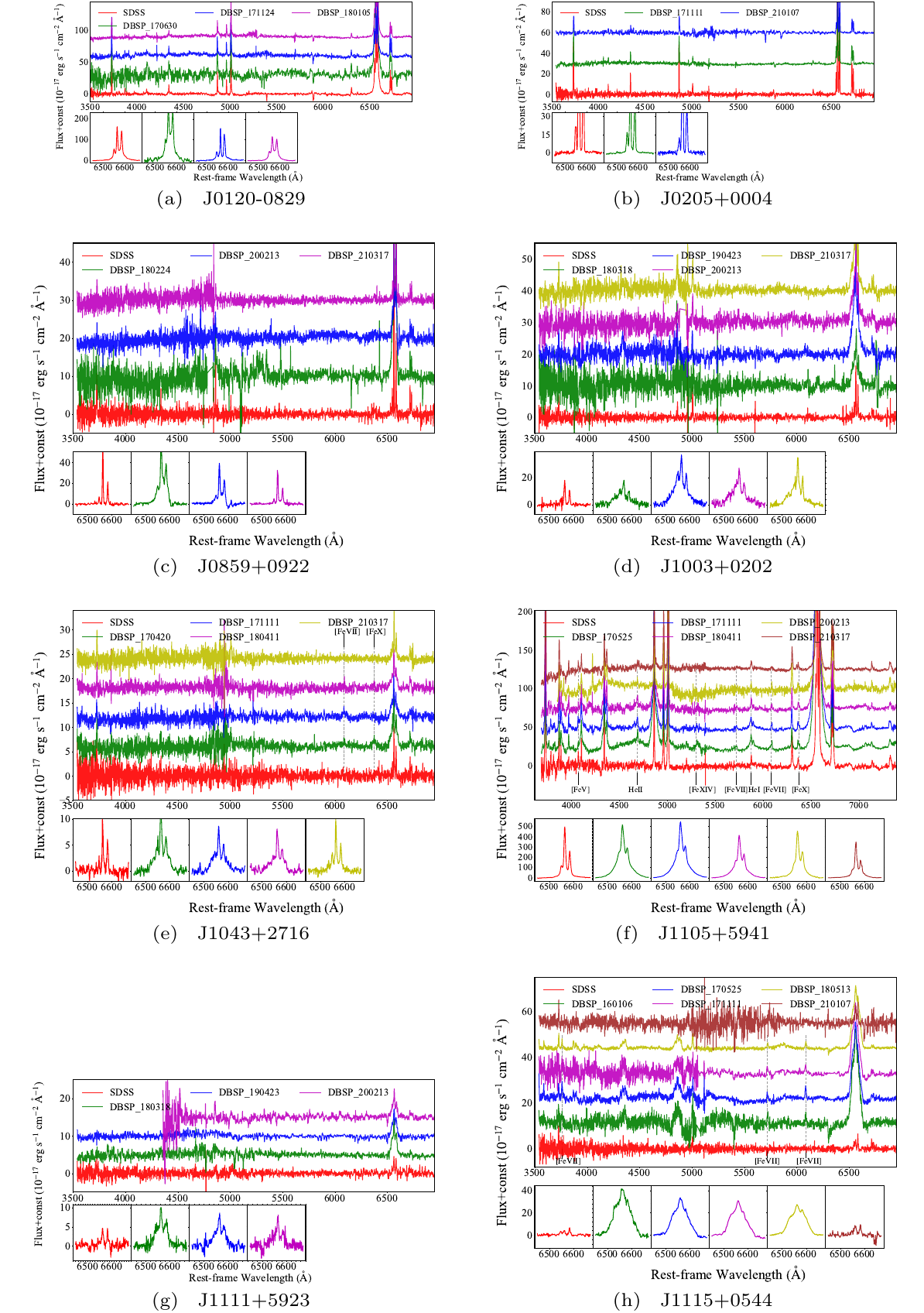}}
\end{minipage}
\caption{Spectral evolution of variable sources. For the source J1111+5923, the original flux calibration of Lick-180318 was not good, and we took the recalibrated one by making the narrow $\rm H\alpha$ flux the same as that of DBSP-200213 because they have similar observation conditions.\label{SpecEvol1} }
\end{figure*}

\begin{figure*}
\figurenum{6}
\centering
\begin{minipage}{0.9\textwidth}
\centering{\includegraphics[angle=0,width=1.0\linewidth]{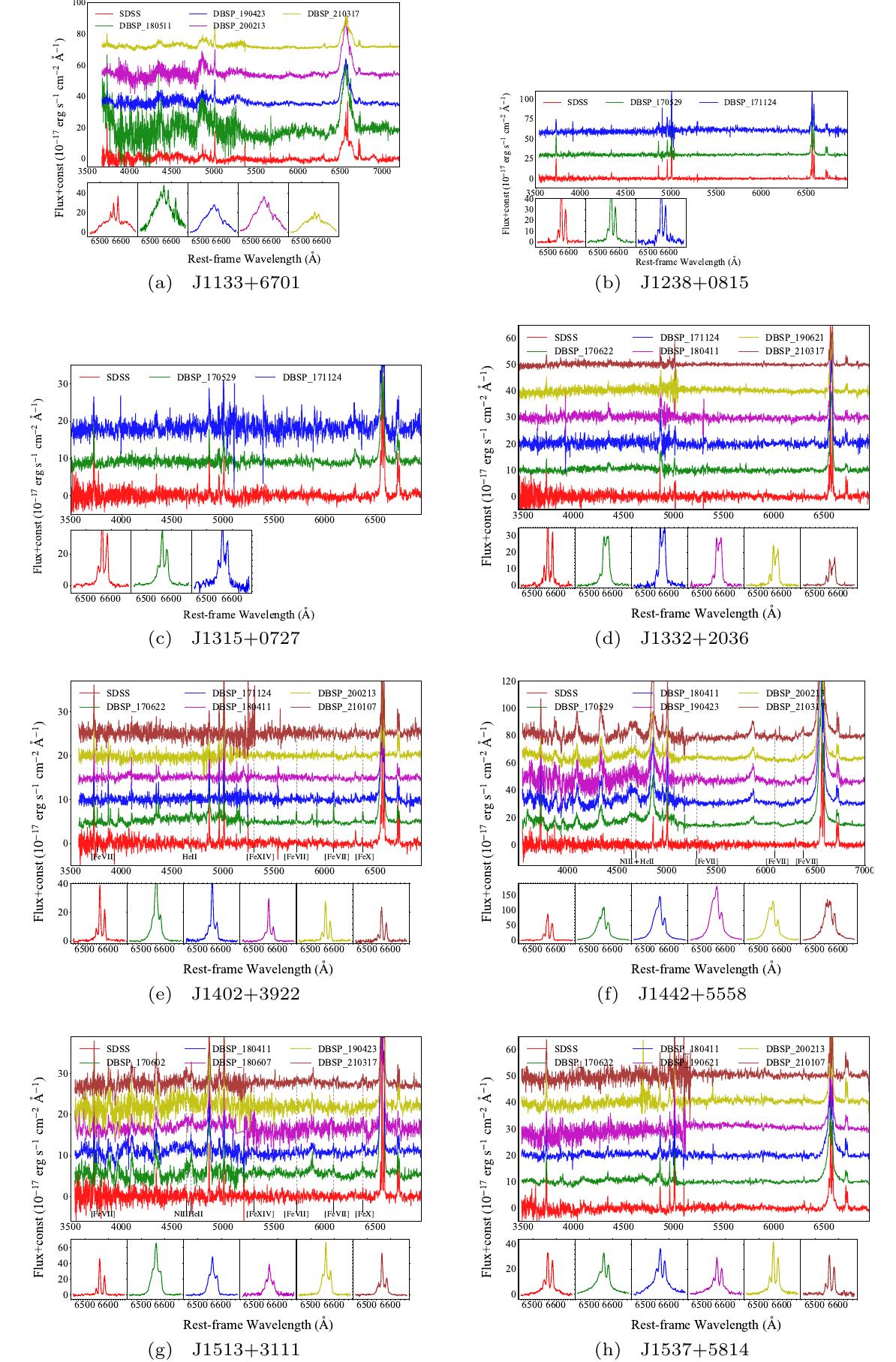}}
\end{minipage}
\caption{Spectral evolution of variable sources. For the source J1133+6701,the recalibration was not applied for DBSP\_180511 because of the bad S/N of its $\rm [O\,\textsc{iii}]\lambda5007$. \label{SpecEvol2}}
\end{figure*}

\begin{figure*}
\figurenum{7}
\centering
\begin{minipage}{1\textwidth}
\centering{\includegraphics[angle=0,width=1.0\linewidth]{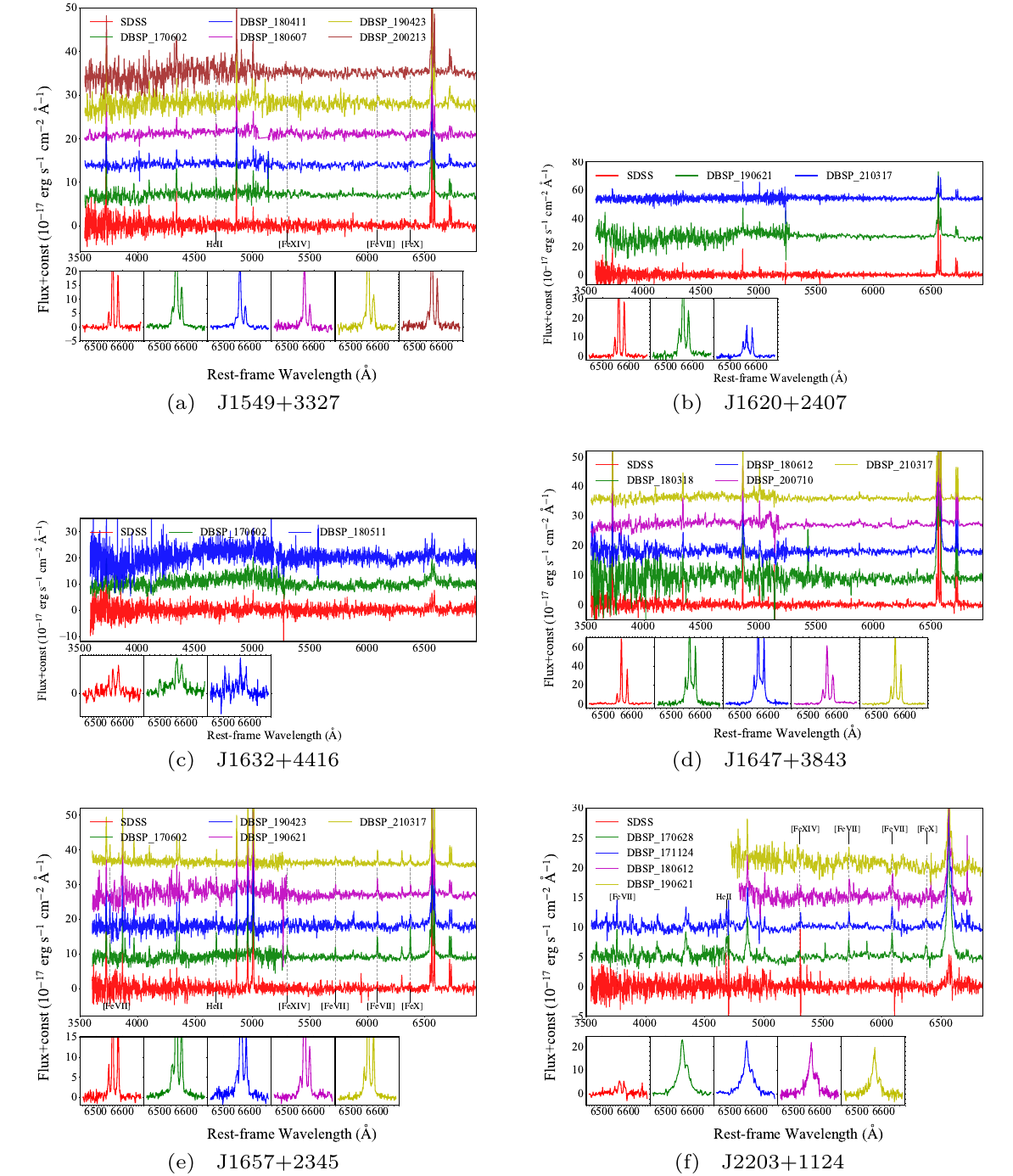}}
\end{minipage}
\caption{Spectral evolution of variable sources. \label{SpecEvol3}}
\end{figure*}

All 22 variable sources' spectral evolution are shown in Figure \ref{SpecEvol1}, \ref{SpecEvol2}, \ref{SpecEvol3} in this appendix. 

\end{appendices}

\clearpage
\twocolumngrid

\end{document}